# Cosmological Dynamics from Modified $f(R)$ Gravity in Einstein Frame

Daris Samart [1]

*Fundamental Physics & Cosmology Research Unit*

*The Tah Poe Academia Institute (TPTP), Department of Physics*

*Naresuan University, Phitsanulok 65000, Thailand*

**Abstract**

In this paper, we investigate and analyze the cosmological dynamics of the universe, with an effect of modified $f(R)$ gravity emerging at cosmological scale. We choose the Einstein frame as a physical frame. We consider phase portraits of the universe at the late time from modified $f(R)$ gravity model. This result gives our universe an acceleration phase expansion without introducing existence of dark energy dominating our universe.

## 1 Introduction

One of the greatest problems in cosmology today is the problem of accelerated expansion of the universe. Fortunately, we can directly obtain various data parameters of the universe from observations. The highly precise observational data from both high presifted supernovae type Ia [1] and the anisotropy power spectrum of CMB [2] exactly tell us that our universe is increasing its rate of expansion to more than that in the past. Almost every cosmologist believes that the universe must be dominated by some mysterious field called dark energy (for interesting recent review see [3]), which drives the accelerated expansion of the universe. An alternative explanation that solves this problem modifies Einstein's gravity theory (i.e. general relativity, or GR for short) to account for the accelerated expansion (for general review see [4]). Carroll et al. [5] proposed a modified model of gravity by adding an inverse power of Ricci scalar into the Einstein-Hilbert action. It resulted in a number of deviations from GR at small curvature and cosmological scale, in which the universe is observed to be expanding with acceleration. The model gives us a de Sitter- and an anti-de Sitter-space solutions in vacuum case, which provide us with a purely gravitation-driven universe, whose expansion is accelerated. It is therefore in good agreement with current observational data. This model is approached by standard variational principle [5], i.e. the action is varied with respect to the metric only. The result of which is the fourth order differential equation. Unfortunately, the model suffers instabilities problem [6], and has been ruled out, despite its equivalence to scalar-tensor theory [7], by a solar system experiment considering a gravitational bound state. The problem seemed to be resolved after Vollick [8] used Palatini variational approach that treated the metic and affine connection as independent variables in order to derive the field equation, which is a second order differential equation [11]; and later, Nojiri and Odintsov [29] solved the same problem by using metric formalisms in which $R^2$ is added into the action of the $1/R$ model. However, Ref. [12] commented that this model should have a conflict in itself, because it was also suitable for a particle experiment at small energy scale. But Ref. [13] pointed out that the method which was discussed by Ref. [12] was just a mathematical equivalence, not a physical one (for extended discussion see [10, 11]). After all, the interesting of this model is strengthened by the fact that it can be derived from string/M-theory, as shown by Ref. [14]. Several authors have worked on various aspects of the model and extended it [16, 21, 22, 29]. However, almost all of these works used the model to find modified versions of Friedmann equation [8, 21, 24, 29] to fit with or be constrained by such observational data as SNa Type Ia, CMB anisotropy. Only a handful of works [18, 19] did investigate or analyze the dynamical system of this model. Ref. [18] considered the phase portraits of $\dot{H}$ vs $H$ (time derivative of Hubble parameter and itself) in Jordan frame, whereas [19] considered between $\dot{R}$ vs $R$ (time derivative of Ricci scalar and itself). We will consider cosmological dynamics of modified gravity with variables identical to that of Ref. [19], but the metric variation formalism in phase portrait of Einstein frame is used instead. In our scenario, the Einstein frame is the physical frame, which gives us self-gravity of the scalar field's effective potential $V(\phi)$ (i.e. $\phi$ in Einstein frame representative form of Ricci curvature in Jordan frame) [17, 23, 26]. This paper is organized as follows: the conformal transformation between Jordan frame and Einstein frame is briefly reviewed in section 2, plus the use of Einstein frame to consider $f(R)$ model of gravity, in which the modified Friedmann equation is found by following the process of [30]; next the $f(R)$ gravity model is investigated and analyzed in section 3 by using cosmological variables obtained from the Friedmann equation. Finally we conclude with problems and the future direction for this paper.

---

[1] E-mail address: *jod_daris@yahoo.com*

## 2   Conformal Transformation Approach to the Einstein Frame and Friedman Equation

In this section, we review the conformal transformation method following [24, 25, 26]; see [20] for the standard derivation. The action of $f(R)$ gravity in Jordan frame with matter field can be written in the form

$$S_J[g_{ab}] = \frac{1}{2\kappa^2} \int f(R) \sqrt{-g}\, d^4x + S_M \,, \tag{1}$$

where $\kappa^2 = 8\pi G$, and $S_M$ is the action of matter field. Ricci scalar is defined by $R = g^{ab}R_{ab}$, $R_{ab} = R^c{}_{acb}$, and the curvature tensor is defined by

$$R^d{}_{abc} = \partial_c \Gamma^d{}_{ab} - \partial_b \Gamma^d{}_{ac} + \Gamma^e{}_{ab}\Gamma^d{}_{ce} - \Gamma^e{}_{ac}\Gamma^d{}_{be}. \tag{2}$$

Varying this action with respect to the metric tensor $g^{ab}$, we get

$$\frac{\delta S_J}{\delta g^{ab}} = f(R)' R_{ab} - \frac{1}{2}g_{ab}f(R) - \nabla_a \nabla_b f'(R) + g_{ab} \nabla_c \nabla^c f'(R) = \kappa^2 T_{ab}, \tag{3}$$

where $f'(R) \equiv df(R)/dR$, and $T_{ab} \equiv \delta S_M/\delta g^{ab}$ is the energy-momentum tensor. Note that the field equation derived from (3) in Jordan frame gives us a fourth order differential equation, which is too complicated to solve [5, 18]. Magnano and Sokolowski [26] was argued the $f(R)$ theories in vacuum case. There are is the existence of the Einstein frame that $\phi$ (we will define it below) is minimally coupling to the metric of the physical frame. We will analyze the late time universe where we assume universe is vacuum. So we choose the Einstein frame as a physical frame. The metric formalisms seem to be equivalent to Palatini formalisms, i.e. $g_{ab}$ is compatible with $\nabla_a$, therefore an affine connection is necessarily equal to the Christoffel symbol in general relativity ($f(R) = R$) [20]. But both variation methods are obviously different in the nonlinear function of the Ricci scalar ($f(R)$ is a higher term which metric formalisms give rise to a fourth order equation but Palatini formalisms give rise to a second order equation [9, 11]), and we obtain the conformal transformation between two frames as [2]

$$^E g_{ab} = e^\phi g_{ab}, \tag{4}$$

where we introduce $\phi$ as a new scalar field defined by $\phi \equiv \ln f'(R)$ [15, 23]. We can rewrite the action in (1), which is dynamically equivalent to the Jordan-Helmholtz action without matter field [14, 26, 27, 28], as

$$S_{JH} = \frac{1}{2\kappa^2} \int \left( f(\sigma) + \chi(R - \sigma) \right) \sqrt{-g}\, d^4x, \tag{5}$$

where $\chi$ and $\sigma$ are auxiliary fields. After varying (5) with respect to $\chi$, we get

$$\sigma = R, \tag{6}$$

and after varying (5) with respect to $\sigma$ again,

$$\chi = f'(\sigma). \tag{7}$$

$\chi$ is eliminated by using (7), hence

$$S_{JH} = \frac{1}{2\kappa^2} \int \left( f(\sigma) + f'(\sigma)(R - \sigma) \right) \sqrt{-g}\, d^4x. \tag{8}$$

---

[2] It's convenient to obtain the conformal factor $e^\phi$ by using Palatini formalisms where the action is varied with respect to $\Gamma^c{}_{ab}$.

Note that (8) is in the form of the scalar-tensor gravitational action for $f(R)$ gravity (1) in Jordan frame [28]. We can rewrite the action (1) in Einstein frame by applying conformal transformation to (4), (i.e. $g_{ab} \to e^{\phi} g_{ab}$) [14, 26],

$$S_E[{}^E g_{ab}] = \frac{1}{2\kappa^2} \int \left( {}^E R - \frac{3}{2} {}^E g^{ab} \widetilde{\nabla}_a \phi \widetilde{\nabla}_b \phi - V(\phi) \right) \sqrt{-{}^E g}\, d^4 x, \qquad (9)$$

where $\widetilde{\nabla}_a$ is the covariant derivative compatible with ${}^E g_{ab}$, and $V(\phi)$ is the effective potential which depends on a choice of $f(R)$ models. The effective potential is defined by[3]

$$V = \frac{R f'(R) - f(R)}{f'(R)^2}. \qquad (10)$$

Varying the action (9) with respect to ${}^E g^{ab}$, we get

$${}^E R_{ab} - \frac{1}{2} {}^E g_{ab} {}^E R = 3 \widetilde{\nabla}_a \phi \widetilde{\nabla}_b \phi - \frac{1}{2} {}^E g_{ab} \left( \frac{3}{2} {}^E g^{ab} \widetilde{\nabla}_a \phi \widetilde{\nabla}_b \phi + V(\phi) \right), \qquad (11)$$

whilst varying it with respect to $\phi$, we obtian

$$\widetilde{\nabla}_a \widetilde{\nabla}^a \phi + \frac{V_\phi}{3} = 0, \qquad (12)$$

where $V_\phi \equiv dV/d\phi$. Now the present universe is considered based on FRW metric, which, in spatially flat space, has the form

$$ds^2 = -dt^2 + a^2(t)(dx^2 + dy^2 + dz^2), \qquad (13)$$

where $a(t)$ is a scale factor of the universe.

Now we will calculate the modified Friedmann equation in Einstein frame from the line element (13) and the $(0,0)$ component of the field equation (11). The Friedmann equation then becomes

$$H^2 = \frac{1}{4} \dot{\phi}^2 + \frac{V(\phi)}{6}, \qquad (14)$$

where $H \equiv \dot{a}/a$ is the Hubble parameter, $\dot{a} \equiv da/dt$, and (15) is used to obtain (14). We also directly obtain the equation of motion for the scalar field $\phi$ from (12) in Einstein frame metric, i.e.

$$\ddot{\phi} + 3H \dot{\phi} + \frac{V_\phi}{3} = 0. \qquad (15)$$

In the next section, we will consider cosmological dynamics in phase portraits of some particular types of $f(R)$ gravity model.

---

[3]Our effective potential form is equivalence to the Palatini formulation [17].

# 3 Cosmological Dynamics of $f(R)$ Gravity

In this section, we will consider cosmological evolution in the phase portrait of Einstein frame, with self gravitating scalar field, to investigate and analyze the acceleration phase for some types of $f(R)$ gravity models. We follow the analysis of [30].
Ref. [29, 31] was demonstrated mechanism of inflation start/end and acceleration start via the potential of curvature evolution. Therefore, we can analogue this mechanism by manifest the scalar field dominates the late time universe in the Einstein frame (the curvature is very small). The acceleration can happen if the field at late time is moving very slowly (i.e. the curvature decrease very slow). Therefore, we can use the slow-roll approximation [33] in this phase [30], i.e. $\ddot{\phi} \approx 0$ and $\dot{\phi}^2 \ll V$. We use the slow-roll approximation only to find the late time trajectory in the phase space, both the Friedman equation and and the equation of motion of the scalar field are

$$H^2 \simeq \frac{V}{6}, \qquad (16)$$

$$3H\dot{\phi} \simeq -\frac{V_\phi}{3}. \qquad (17)$$

By applying the slow-roll approximation condition to both the Friedmann equation in (16), and the equation of motion for the scalar field in (17), then we directly get the late time trajectory in the phase space portrait as

$$\dot{\phi} \simeq \frac{V_\phi}{3\sqrt{6V}}. \qquad (18)$$

Now we derive the accelerating condition of the universe in $f(R)$ gravity model. The accelerating condition of the expansion is

$$\frac{\ddot{a}}{a} = \dot{H} + H^2 > 0, \qquad (19)$$

which implies that

$$-\frac{\dot{H}}{H^2} < 1. \qquad (20)$$

We obtained $\dot{H}$ directly by differentiating $H^2$ in (16), and together with using (17), we then obtain

$$\dot{H} = -\frac{1}{18}\frac{V_\phi^2}{V}. \qquad (21)$$

By using (16) and (21) substituting in (20), it's becomes

$$\left(\frac{V_\phi}{V}\right)^2 < 3. \qquad (22)$$

These equations are the accelerating conditions of expansion. In the previous section, we have derived the Friedmann equation in Einstein frame metric. We will derive the dynamical equation of $\phi$ and $\dot{\phi}$. For obtain new value of $\dot{H}$ with out slow-roll approximation by using (14) and (15)

$$\dot{H} = -\frac{3}{4}\dot{\phi}^2. \tag{23}$$

We define a new variables as

$$X = \phi, \quad Y = \dot{\phi}. \tag{24}$$

Using new variables in (24) substitute into (23) and (15), we obtain autonomous system as

$$\begin{aligned}
\dot{H}(X,Y) &= -\frac{3}{4}Y^2 \\
\dot{X} &= Y \\
\dot{Y} &= -3H(X,Y)Y - \frac{1}{3}\frac{dV(X)}{dX}.
\end{aligned} \tag{25}$$

The function $H(X,Y)$ is given by

$$H(X,Y) = \sqrt{\frac{1}{4}Y^2 + \frac{1}{6}V(X)}. \tag{26}$$

These equations will be used for the analysis in the phase portrait of the $f(R)$ modified gravity. Next, we will investigate and analyze the phase space portraits for some types of the effective potential $V(\phi)$, which depends on the choice of $f(R)$ models of gravity. By using the autonomous system of equations in (25). We also show the acceleration condition curve which divides acceleration and non-acceleration regions in the phase space portraits.

## 3.1 $f(R) = R - \mu^{2(n+1)}/R^n$ model

This model was proposed by [5] (for $n = 1$) at the cosmological scale when curvature is very small at the late-time universe. Our gravitational Lagrangian density ($n = 1$) in Jordan frame has

$$f(R) = R - \mu^4/R, \tag{27}$$

where $\mu^4$ is a positive-valued parameter. The effective potential $V(\phi)$ in Einstein frame depends on the choice of $f(R)$. At the late time, $R$ is very small, we get

$$e^\phi \equiv f'(R) = 1 + \frac{\mu^4}{R^2} \simeq \frac{\mu^4}{R^2}. \tag{28}$$

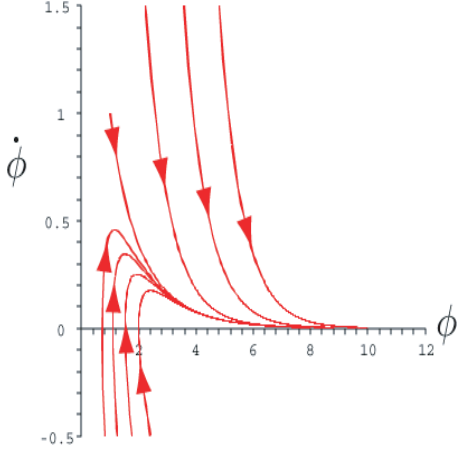

Figure 1: The phase portrait of $\dot\phi$ vs $\phi$ in the modified $1/R$ gravity model when $\mu = 1$. We note that everywhere in phase space portrait is an acceleration region following the condition in (32).

Then, the effective potential in the Einstein frame of the model $f(R) = R - \mu^4/R$ gravity at the late time by using in (28) substituted into (10) in [29]

$$V(\phi) \simeq \frac{2}{\mu^2} e^{-\frac{3}{2}\phi}. \tag{29}$$

By plugging the effective potential (29) into (18), then the late time trajectory is

$$\dot\phi = \frac{1}{2\sqrt{3}\mu} e^{-\frac{3}{4}\phi}. \tag{30}$$

The effective potential of our late time approximation looks like the power law inflation, which gives scale factor evolution proportional to the power of time, $a(t) \sim t^p$. This proportionality comes from the potential [4] $V \sim \exp\left(-\frac{2}{p}\phi\right)$ with $p = \frac{4}{3}$ in our case [33, 34], such that

$$a \sim t^{4/3}. \tag{31}$$

We note that potential increases when curvature is minimum and therefore the universe begins to have the power law expansion and substitute (29) into (22), then accelerating condition is

$$\left(\frac{\frac{d}{d\phi}\left(\frac{2}{\mu^2} e^{-\frac{3}{2}\phi}\right)}{\frac{2}{\mu^2} e^{-\frac{3}{2}\phi}}\right)^2 < 3. \tag{32}$$

---

[4] Our convention compared with standard power law formalisms is $\phi \equiv \sqrt{\frac{p}{2}}\varphi$ with potential taking the form $V \sim \exp(-\sqrt{\frac{2}{p}}\varphi)$.

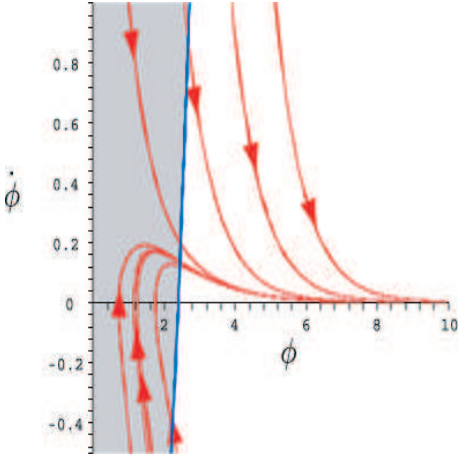

Figure 2: The phase portrait of $\dot{\phi}$ vs $\phi$ in the modified $\ln R$ gravity model, when $\alpha = 1$, the acceleration region divided by acceleration condition (blue line) where acceleration region is bright region and shaded region is no acceleration region.

The above condition obviously shows that everywhere in the phase space portrait of $\dot{\phi}$ vs $\phi$ in $f(R) = R - \mu^4/R$ gravity model can be an acceleration region, i.e. wherever any attractor occurs in phase space, it can always generate the accelerated expansion in both present and late time phase and the phase plot will occur attractor along the line of the late time trajectory curve (30); See Figure 1.

## 3.2 $f(R) = R + \alpha \ln\left(R/\gamma^2\right)$ model

This model was proposed by [31] to solve the problem of instabilities in the fourth order differential equation from CDTT's model in metric formalisms [5]. It also provides accelerating phase at late time. (For more detail of Palatini formalisms of this model, see [32]). The gravitational Lagrangian density of this model in Jordan frame is

$$f(R) = R + \alpha \ln(R/\gamma^2), \tag{33}$$

where $\alpha$ and $\gamma$ are positive-valued parameters. As in the previous subsection, at the late time $R$ is very small, we have

$$e^\phi \equiv f'(R) = 1 + \frac{\alpha}{R} \simeq \frac{\alpha}{R} \tag{34}$$

then we obtain the effective potential in the Einstein frame of the model $f(R) = R + \alpha \ln(R/\gamma^2)$ gravity at the late time by plugging (34) into (10), as in [31]

$$V(\phi) \simeq \alpha e^{-2\phi} \phi. \tag{35}$$

The late time trajectory is

$$\dot{\phi} = \frac{\alpha e^{-\phi}(1 - 2\phi)}{3\sqrt{6\phi}}. \tag{36}$$

This effective potential of our late time approximations that looks like generalized exponential potential was originally examined by J. D. Barrow [35], and extended investigation was done by [36], i.e. the potential takes the form $V \sim \phi^m \exp(-\lambda \phi^n)$. For our case, $m = n = 1$ which gives scale factor evolution that is proportional to the power of time as [31]

$$a(t) \sim t^{\frac{3}{4}}. \tag{37}$$

We performed calculations following the previous subsection by plugging (35) into (22), the accelerating condition is then

$$\phi^2 - 2\phi - 1 < 0. \tag{38}$$

The above condition manifests itself as a curve (blue line) dividing the acceleration region (bright region) from the no acceleration region (shaded region). We note that from Figure 2, an attractor occurs in the phase plot in acceleration region, therefore, the universe's expansion will be accelerated in $f(R) = R + \alpha \ln(R/\gamma^2)$ gravity model, and the attractor occurs in the phase plot along the line of the late time trajectory curve in (36); See Figure 2.

## 4 Conclusions

We have investigated the modified $f(R)$ gravity models, namely $1/R$ and $\ln R$ gravity models. Both models give the late time attractors. These results mean that our universe has its expansion accelerated. The evolution when scalar field is at some given point on the potential has to be independent of any initial conditions. The attractor of $1/R$ gravity can drive the universe accelerating its expansion, because the condition of acceleration allows everywhere in phase space portrait to be an acceleration region. In the $\ln R$ gravity, an attractor occurs in the acceleration region according to the acceleration condition. The universe can then accelerate its expansion in this model. It is surprising that our analysis in Einstein frame gives similar result as in [37] which chose Jordan frame as a physical frame. They showed that $f(R)$ model gravity has a late time attractor on a phase space portrait of $\dot{H}$ vs $H$. Both Jordan and Einstein frames are analyzed similarly for $f(R)$ gravity model. Our investigation and analysis give well behavior in cosmological dynamics at late time. The $f(R)$ gravity theories also provide inflation at early time by adding higher curvature in the gravitational action [29, 38]. These results are confirmed by observational data in several works that we have discussed above. It also passes the solar system experimental tests, and gives no stabilities problem [29, 31, 39]. Hopefully, the modified $f(R)$ gravity is a viable candidate to dark energy. Although our analysis in Einstein frame gives late time acceleration phase, Carroll et al.[5] originally analyzed universe in Jordan frame. They considered evolution in Einstein frame and transformed it back to Jordan frame as the physical frame. Because, in Einstein frame, the energy density $\rho$ evolution with scale factor $\rho \sim a^{-3}$ does not hold in the ordinary matter dominated case [5], which is the problem that will be investigated in the future work. However, we cannot tell whether the Jordan frame or Einstein frame is a physical frame. For general consideration and mathematical equivalence (via conformal transformations) between Eintein, Jordan and even ideal fluid frames of $f(R)$ gravity was given in detail in [10].

**Note Added** During the preparation of this work. Modified gravity which seems to be quite successful as dark energy was criticized by recent papers from L. Amendola, D. Polarski and S. Tsujikawa appeared on arXiv [40] because some simplest version of it like $1/R^n$ gravity are not able to describe at once and matter dominated phase, then transition from matter dominated phase to acceleration, then acceleration. However, some counterexamples where such sequence of universe epochs is possible in more complicated class of $f(R)$ gravity ware suggested in S. Capozziello, S. Nojiri, S. D. Odintsov and A. Troisi [41].

**Acknowledgement** I would like to thank Burin Gumjudpai, my supervisor, M. Sami and S. Tsujikawa for a useful discussion, critical comments; Prof. S. D. Odintsov for useful discussion and references; Itzadah Thongkool for a helpful discussion; Narit Pidokrajt for critical reading this manuscript; Kiattisak Thepsuriya for English proofreading; Sarayut Pantian for his help and support in numerical results; and Chakkrit Kaeonikhom for nice graphics.

---

[5] I would like to thank Shinji Tsujikawa for pointing out this problem and for kind discussions.